\documentclass[aps,prx,twocolumn,final,letterpaper]{revtex4}

\usepackage{appendix}
\usepackage{graphicx}   
\usepackage{import}
\usepackage{comment}
\usepackage{epstopdf}
\usepackage{bm}
\usepackage{amssymb}
\usepackage{quotes}
\usepackage{braket}
\usepackage{transparent}
\usepackage{dcolumn}
\usepackage{multirow}
\usepackage{cancel} 
\usepackage{mdframed}
\usepackage{color}
\usepackage{bm}
\usepackage{dsfont}
\usepackage{slashed}
\usepackage{enumitem}
\usepackage{amsmath} 
\begin{document}
\rmfamily

\title{Multimode amplitude squeezing through cascaded nonlinear optical processes}

\author{Sahil Pontula$^{1,2,3}$, Yannick Salamin$^{1,3}$, Charles Roques-Carmes$^{3,4}$, Marin Solja\v{c}i\'{c}$^{1,3}$}
\affiliation{$^1$ Department of Physics, MIT, Cambridge, MA 02139, United States. \\
$^2$Department of Electrical Engineering and Computer Science, MIT, Cambridge, MA 02139, United States.\\
$^3$ Research Laboratory of Electronics, MIT, Cambridge, MA 02139,  United States.\\
$^4$ E. L. Ginzton Laboratories, Stanford University, 348 Via Pueblo, Stanford, CA USA.}

\begin{abstract} 

Multimode squeezed light is enticing for several applications, from squeezed frequency combs for spectroscopy to signal multiplexing in optical computing. To generate squeezing in multiple frequency modes, optical parametric oscillators have been vital in realizing multimode squeezed vacuum states through second-order nonlinear processes. However, most work has focused on generating multimode squeezed vacua and squeezing in mode superpositions (supermodes). Bright squeezing in multiple discrete frequency modes, if realized, could unlock novel applications in quantum-enhanced spectroscopy and optical quantum computing. Here, we show how $Q$ factor engineering of a multimode nonlinear cavity with cascaded three wave mixing processes creates strong, spectrally tunable single mode output amplitude noise squeezing over 10 dB below the shot noise limit. In addition, we demonstrate squeezing for multiple discrete frequency modes above threshold. This bright squeezing arises from enhancement of the (noiseless) nonlinear rate relative to decay rates in the system due to the cascaded generation of photons in a single idler ``bath'' mode. A natural consequence of the strong nonlinear coupling in our system is the creation of an effective cavity in the synthetic frequency dimension that sustains Bloch oscillations in the modal energy distribution. Bloch mode engineering could provide an opportunity to better control nonlinear energy flow in the synthetic frequency dimension, with exciting applications in quantum random walks and topological photonics. Lastly, we show evidence of long-range correlations in amplitude noise between discrete frequency modes, pointing towards the potential of long-range entanglement in a synthetic frequency dimension.

\end{abstract}

\maketitle

\section{Introduction}
\label{sec:intro}

Quantum states of light prepared using nonlinear parametric processes have attracted significant interest for applications in precision measurement and quantum technologies through noise squeezing and entanglement properties \cite{aasi2013enhanced,madsen2022quantum, de2014full, mcculler2020frequency}. Second order nonlinear processes have emerged as a key platform to generate quantum states of light by processes including second harmonic generation and parametric downconversion in optical parametric oscillators (OPOs), leading to numerous theoretical proposals and experimental realizations of entangled single photon pairs, single mode squeezing, squeezed supermodes (including two-mode squeezing), and broadband quantum frequency combs \cite{louisell1961quantum, wu1986generation, vahlbruch2008observation, roslund2014wavelength, dutt2015chip, ra2020non, zhao2020near, vaidya2020broadband, yang2021squeezed, guidry2023multimode, presutti2024highly, stokowski2024integrated, jankowski2024ultrafast}. However, most of these works have focused on single mode and multimode squeezed vacua (as generated, for example, by OPOs pumped below threshold). Nondegenerate OPOs operated above threshold have also been investigated for amplitude squeezing. In many works, twin beam squeezing is often considered, given the strong correlation between intensity noise in signal and idler modes from a single parametric downconversion process. Single beam squeezing is also possible, considering the photon number filtering that twin beam correlations induce \cite{fabre1989noise, mertz1990observation, furst2011quantum}. However, the output amplitude noise squeezing for a single nondegenerate downconversion process is theoretically limited to 3 dB below the shot noise limit \cite{fabre1989noise}. This squeezing is also generally limited to narrow spectral ranges depending on the nonlinear crystal used and its phase matching conditions. Achieving tunable single mode and multimode ``bright'' amplitude squeezing in systems with multiple frequency modes remains unexplored, despite exciting potential for quantum optical information multiplexing and bright squeezed frequency combs for spectroscopy applications \cite{fabre2020modes}.



Here, we explore a novel scheme for amplitude squeezing within a multimode cavity with second-order nonlinearity that employs cascaded parametric amplification mediated by a common terahertz ``bath'' mode to create an infrared frequency comb with terahertz mode spacing \cite{burgess2009difference, ravi2016cascaded, hemmer2018cascaded}. We demonstrate that by strategically engineering the cavity's $Q$ factor profile, we can manipulate the nonlinear energy flow through frequency space, thereby significantly shaping both mean field and noise properties. Our method works by creating a high $Q$ factor cavity in frequency space that traps nonlinear energy flow within a finite (and controllable) span of discrete frequency modes \cite{hu2022mirror}. This enables very strong nonlinear coupling between nearest-neighbor frequency modes that can exceed decay rates in the system. This can lead to excitation of counter-propagating Bloch modes in the ``frequency cavity,'' whose interference is manifested in a standing wave distribution of steady state modal energies.
By increasing the outcoupling (lowering the loaded $Q$ factor) for one or more selected discrete frequency modes, we show that the frequency cavity supports simultaneous output amplitude squeezing in these mode(s) over a $>100$ MHz bandwidth. Strong squeezing emerges because of an enhancement in the noiseless nonlinear coupling over dissipation. Finally, we describe how the strong nonlinear interactions in our system create strong long-range correlations in amplitude noise that may suggest existence of long-range entanglement in a synthetic frequency dimension. Our study of quantum noise through cascaded nonlinear interactions suggests many exciting possibilities, including bright squeezed frequency combs, trapped states and solitons in the synthetic frequency dimension, tunable quantum walks, and much more \cite{hu2022mirror}. 

The rest of this paper is structured as follows. In Sec. \ref{sec:theory}, we introduce our system and describe how its mean field dynamics and quantum noise behavior can be calculated. Included in this section is a discussion of the ``frequency cavity'' that our system realizes, which supports counter-propagating resonant Bloch modes that cause interference patterns in the modal energy distribution. In Sec. \ref{sec:intra}, we describe intracavity mean field dynamics, noise, and the presence of relaxation oscillations due to strong multimode nonlinear coupling. In Sec. \ref{sec:output}, we show tunable single mode and multimode squeezing in output amplitude noise, describing the limitations and conditions for generating strong squeezing. Lastly, in Sec. \ref{sec:correl}, we show how the coupling of the infrared frequency comb to a common terahertz bath mode allows strong long-range correlations in frequency space.


\section{Theory}
\label{sec:theory}

\begin{figure*}[t]
    \includegraphics[scale=0.8]{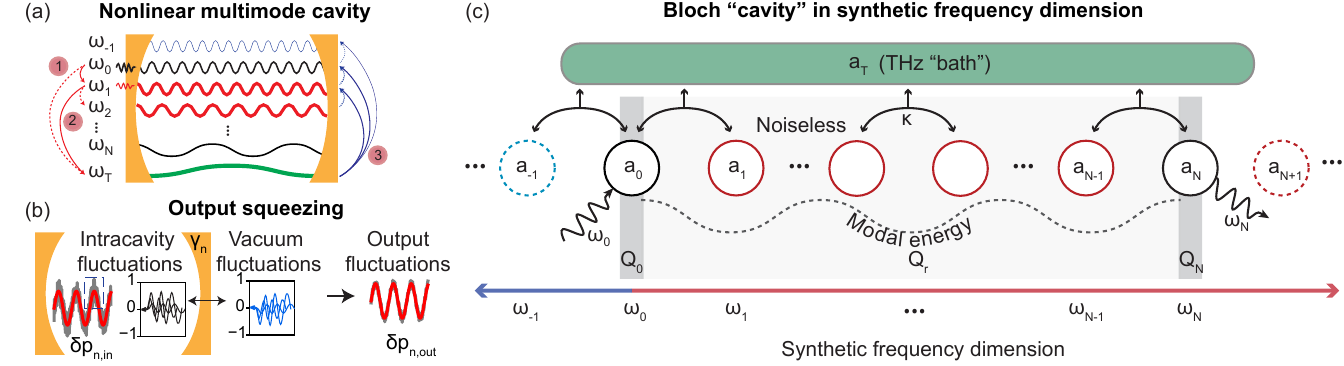}
    \caption{\textbf{Squeezing in a multimode cavity with THz-mediated cascaded three wave mixing.} (a) Cascading infrared (IR) orders are resonant in a multimode cavity and undergo three wave mixing (TWM) mediated by a terahertz (THz) mode, creating a frequency comb (red) with modes separated by the THz frequency $\omega_T$ (green). The cascade starts with a single TWM process wherein a pump photon at $\omega_0$ amplifies a seed photon at $\omega_1$ (solid line) and simultaneously creates an idler photon (THz, dashed line) (1). Subsequently, the amplified mode at $\omega_1$ initiates cascading downconversion processes, now seeded by the THz idler photon (2). Concomitantly, THz photons can also initiate upconversion processes that repopulate the IR orders (3). By shaping the $Q$ factor distribution of the cavity (e.g., through a frequency-dependent coupler), the modes blueshifted relative to the pump frequency $\omega_0$ can be suppressed, biasing downconversions that create THz photons. Through parametric squeezing enabled by the strong nonlinear rates, the multimode cavity can create above-threshold output squeezing in frequency mode(s) that are separated from the coherent pump mode by multiple idler photons. (b) Shown for a single mode, the output squeezing emerges due to destructive interference between the intracavity fluctuations and vacuum shot noise on the output facet of the cavity outcoupling mirror. (c) Strong squeezing requires strong nonlinear energy flow, which creates a kind of nonlinear tight binding system in frequency space. The system is bounded by low $Q$ modes at frequencies $\omega_{0,N}$, resulting in a frequency space cavity (modes within the cavity generally have high $Q$ factors). Excitation of counter-propagating Bloch modes in this cavity creates an interference pattern that is observable in the modal energy distribution.}
    \label{fig:fig1}
\end{figure*}

At the heart of our concept is the novel $Q$ engineering of a multimode nonlinear cavity \cite{salamin2021overcoming}, as illustrated in Fig. \ref{fig:fig1}a. Our system comprises a comb of infrared (IR) cavity modes, each mode coupled to its nearest neighbors via a bath mode, here specifically a terahertz (THz) frequency mode (whose small frequency allows for many modes in a small IR span). The process begins with nonlinear three wave mixing (TWM) where a pump photon ($\omega_0$) simultaneously amplifies a seed photon ($\omega_1$) and generates a new photon in the idler ``bath'' mode ($\omega_T$). Subsequently, cascading steps are initiated by the nonlinear interaction of the bath mode (THz) with the IR modes, resulting in multiple equally spaced modes. This produces a frequency comb with spacing given by $\omega_T$. By properly engineering the $Q$ factors of the cavity modes, one can favor the three wave downconversion processes that create THz idler photons. This enhances the rate of nonlinear energy flow in the cavity and, due to its noiseless nature, can result in squeezing when the nonlinear rate surpasses dissipation rates in the system.

We design the multimode cavity such that, through phase matching constraints, only TWMs of the form $\omega_{n-1}\leftrightarrow \omega_{n}+\omega_T$ are supported (here, $\omega_k<\omega_n$ when $k>n$). The decay rates for modes of all other frequencies are assumed much faster than the relevant timescales in this system, so we restrict our attention to the system specified by the coupled IR modes and the THz mode.  

We simulate the mean field and noise properties of our system by using the Heisenberg-Langevin equations of motion for the mode field (annihilation) operators (see Appendix for details), which read
\begin{align}
    \begin{split}
        \dot a_T &= \kappa\sum_n a^\dagger_na_{n-1} - \gamma_T a_T + \sqrt{2\gamma_T}s_T \\
        \dot a_n &= \kappa\left(a_T^\dagger a_{n-1} - a_{n+1} a_T\right) - \gamma_n a_n + \sqrt{2\gamma_n}s_n,
    \end{split}
\end{align}
where $a_{n}$ are field operators that determine the photon number $\langle a_{n}^\dagger a_{n} \rangle$, $\kappa\in \mathbb{R}$ has units of s$^{-1}$ and denotes the nonlinear coupling strength (related to the nonlinear susceptibility $\chi^{(2)}$ and assumed frequency-independent), $\gamma_n,\gamma_T$ denote the outcoupling rates for the IR and THz modes, $s_n$ denotes the external fields, and the indexing is such that $n>0$ correspond to redshifted modes relative to the pump at $n=0$. An estimation of typical values of $\kappa$ for realistic experimental settings is provided in the Appendix. Unless specified otherwise, only modes $a_{0,1}$ are pumped, so that only $s_{0,1}$ have nonzero mean. Note that in the Heisenberg-Langevin formalism, the zero-mean terms $\sqrt{2\gamma_n}s_n$ are Langevin forces associated with the outcoupling process (see Appendix). In Sec. \ref{sec:output}, we consider the effect of intrinsic loss, which adds further noise to the system through other Langevin forces (the full equations of motion including intrinsic loss are provided in the Appendix). In our simulations, we numerically solve the Heisenberg-Langevin equations of motion in the mean field domain using backward differentiation \footnote{This allows us to handle numerically stiff systems with strong nonlinearity.} to obtain the steady state mode amplitudes $a_n\equiv \langle a_n\rangle,a_n^* = \langle a_n^\dagger\rangle$ from a vacuum initial state.

Assuming a strong, coherent pump (up to small fluctuations), when the system reaches steady state, we can linearize the equations of motion about the mean values for the fields ($\hat a_n = a_n + \delta\hat a_n$ where here we made explicit the distinction between an operator and complex number) to construct linearized equations of motion for the operators $\delta a_n,\delta a_n^\dagger$. Note that we do in general need to linearize with respect to two degrees of freedom for each mode ($a_n,a_n^\dagger$). However, by picking zero initial conditions for the fields and taking $\kappa\in\mathbb{R}$, the steady states will be real-valued in this model. Then, we can define quadrature operators $p_n = a_n+a^\dagger_n,q_n = -i\left(a_n - a^\dagger_n\right)$, whose fluctuations directly give the amplitude and phase noise of mode $a_n$ (as shown in the Appendix). Because the modal amplitudes are real-valued, $p_n,q_n$ do not couple. We can perform a Fourier transform and derive a system linear in the fluctuations $\delta p_n(\omega)$ that can be arranged in matrix form as $M(\omega)P(\omega) = F(\omega)$, where $P(\omega)=\left[\delta p_0(\omega) \delta p_1(\omega) \cdots \delta p_N(\omega) \delta p_T(\omega)\right]^T$ and $F=\left[F_0(\omega) F_1(\omega) \cdots F_N(\omega) F_T(\omega)\right]^T$ is the Langevin force vector. The zero-mean Langevin forces satisfy $\langle F^\dagger_n F_{n'}\rangle=2\gamma_n\delta_{nn'}$ (see Appendix for details). An explicit expression for the fluctuation matrix $M$ is also provided in the Appendix. The amplitude noise for mode $a_n$ is coupled to the noise of $a_T, a_{n-1}, a_{n+1}$ in the frequency domain according to
\begin{align}
    \delta p_n = \frac{\kappa\left[\delta p_T (a_{n-1}-a_{n+1}) + a_T (\delta p_{n-1} - \delta p_{n+1}) \right] + F_n}{-i\omega+\gamma_n}.
    \label{eq:noise}
\end{align}
From the elements of the inverse fluctuation matrix $M^{-1}$ as well as the Langevin force correlators, the intracavity and output amplitude noise can be computed according to
\begin{align}
\begin{split}
        \langle \delta p^\dagger_{n,\mathrm{in}} \delta p_{n,\mathrm{in}} \rangle &= |M^{-1}_{n,N+1}|^2(2\gamma_T) + \sum_{k=0}^N |M^{-1}_{n,k}|^2(2\gamma_k)\\
    \langle \delta p^\dagger_{n,\mathrm{out}} \delta p_{n,\mathrm{out}} \rangle &= 1 + 2\gamma_n\langle \delta p^\dagger_{n,\mathrm{in}} \delta p_{n,\mathrm{in}} \rangle - 4\gamma_n\mathrm{Re}(M^{-1}_{n,n}),
    \label{eq:inout}
\end{split}
\end{align}
where the output fluctuation amplitude for mode $a_n$ is given by $\delta p_{n,\mathrm{out}}=\sqrt{2\gamma_n}\delta p_{n,\mathrm{in}} - (\delta s_n + \delta s_n^\dagger)$. To compute the squeezing factor, we compare the amplitude noise to the corresponding shot noise limit (SNL) in the absence of any nonlinear processes. Under only driven-dissipative dynamics (and neglecting intrinsic loss), one can show (see Appendix) that $\langle \delta p^\dagger_{n,\mathrm{in}}(\omega) \delta p_{n,\mathrm{in}}(\omega) \rangle=2\gamma_n/(\gamma_n^2+\omega^2)$ and $\langle \delta p^\dagger_{n,\mathrm{out}}(\omega) \delta p_{n,\mathrm{out}}(\omega) \rangle=1$. These represent what we will use as ``reference coherent states'' when analyzing intracavity and output noise in our system. As we will show, enhancement of nonlinear coupling can enable destructive interference between the vacuum field $s_n$ and the intracavity field $p_{n,\mathrm{in}}$, generating output squeezing as shown in Fig. \ref{fig:fig1}b.


\section{Intracavity mean field dynamics and noise}
\subsection{Frequency space cavity}
\label{sec:intra}

Our system is analogous to the implementation of coupled resonator optical waveguides in a synthetic frequency dimension \cite{mookherjea2002coupled}. When the system is truncated by boundary modes in the frequency dimension, it can be thought of as a Fabry-Perot-type cavity in frequency space defined by the finite extent of the high $Q$ factor cascading orders and bounded by lower $Q$ factor frequency ``mirrors.'' The leakiest modes lie outside the frequency cavity. In Fig. \ref{fig:fig1}c, this is shown for a one-sided comb where frequency cavity modes $\omega_{1,...,N-1}$ have high $Q$ factor ($Q_r$) and frequency mirrors at $\omega_{0,N}$ have lower $Q$ factor ($Q_{0,N}$). A natural consequence is that the excitation of modes in this frequency space cavity should manifest in the steady state energy distribution of the frequency modes. We can make this rigorous by considering a Bloch mode analysis, noting that our system is a kind of nonlinear tight-binding model with quasi-discrete translational symmetry (up to boundary conditions at the frequency mirrors) in the synthetic frequency dimension. As a crude approximation, in the case of linear coupling (i.e. assuming $a_T$ is constant) and neglecting dispersion, we have the result that two counter-propagating Bloch waves with wave vectors $k_{\pm}=\pi/2a$ (where $a=\omega_T$ is the lattice constant of the frequency crystal defined by the cascading frequency modes) are excited \cite{hu2022mirror}. (Though beyond the scope of our work, we note that higher order Bloch modes may be excited by using a pump detuned from $\omega_0$. This provides an extra degree of freedom that could allow synthesis of arbitrary states/modal profiles in the frequency dimension.) Interference of the Bloch modes creates a modal energy distribution with quasi-periodicity $2a$ (due to dissipation, modal energy drops further from the pump mode). The magnitude of interference can be tuned via the reflectivity of the frequency mirrors. For example, one can minimize interference by creating an open boundary condition at $\omega_N$. This can be done by ``impedance matching'' mode $a_N$ such that $\gamma_N\approx\kappa a_T$. This results in minimum reflectivity at $\omega_N$. While minimizing interference is advantageous for maximizing efficiency of populating the terahertz idler mode \cite{salamin2021overcoming}, other design methods, which we describe below, are more optimal for maximizing output squeezing. 


We briefly note that the interference state in steady state modal energy naturally translates into interference in the low frequency modal amplitude noise from the linearization procedure. In steady state, $a_n\propto a_{n-1}-a_{n+1}$ and at zero noise frequency, $\delta p_n(0) \propto \delta p_T(0) (a_{n-1}-a_{n+1})$ (the first term in Eq. \ref{eq:noise} is usually dominant since the terahertz mode contains additive noise from all of the TWM processes). Thus, generally, cascading IR modes with higher intracavity intensity are accompanied with higher low-frequency intracavity amplitude noise.

In Fig. \ref{fig:fig2}a, b, we show how a $Q$ engineered multimode cavity that favors frequency downconversions enables strong nonlinear energy flow that creates a frequency comb in modes redshifted relative to the pump mode $a_0$. Fig. \ref{fig:fig2}a shows the temporal dynamics of the modal energy distribution for a multimode cavity with $Q$ factor spectrum given by the first panel of Fig. \ref{fig:fig2}b. Intuitively, energy ``bounces'' back and forth between the two frequency mirrors, eventually creating a steady state modal energy distribution that shows Bloch mode interference (excitation of Bloch modes with wavevectors $k_{\pm}=\pi/2\omega_T$) for the infrared modes lying inside the cavity, as shown in the second panel of Fig. \ref{fig:fig2}b. Only modes trapped within the frequency cavity defined by the frequency mirrors at $\omega_{0,N}$ are appreciably occupied. Of additional note is the high energy of the terahertz idler mode in the steady state, reflecting the ability of our system to generate the terahertz idler mode with high efficiency \cite{salamin2021overcoming}. In the Appendix, we show how one-sided (blueshifted) and two-shifted frequency combs can also be produced in our system with appropriate $Q$ factor shaping.

%
%

\subsection{Intracavity noise}

\begin{figure}[t]
    \includegraphics[scale=0.7]{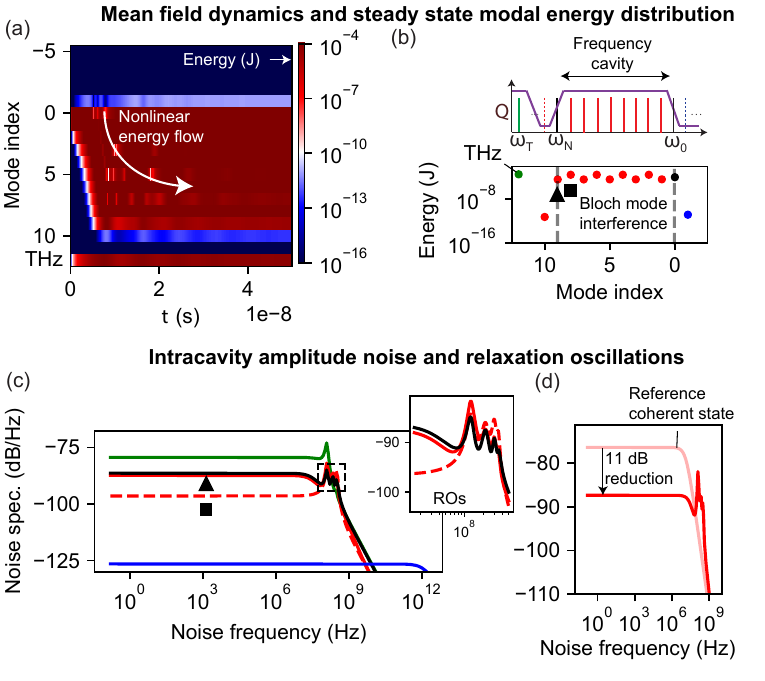}
    \caption{\textbf{Intracavity dynamics and noise due to strong cascaded nonlinear interactions.} (a) $Q$ factor shaping (through the use of frequency-dependent couplers) permits the creation of frequency combs containing only redshifted modes relative to the pump mode $a_0$. The temporal dynamics feature Bloch waves that propagate through frequency space, establishing the steady state interference pattern in intracavity modal energy. (b) $Q$ factors for the different frequency modes and quasi-periodic modal energy distribution in frequency space enabled through cascaded nonlinear interactions. Green denotes the (THz) idler mode $a_T$, red denotes infrared cascading orders $a_{n>0}$, black denotes the pump mode $a_0$, and blue denotes blueshifted modes $a_{n<0}$ (suppressed in the present system). The dashed lines indicate the boundaries of the cavity in the synthetic frequency dimension. (c), (d) Intracavity relative intensity noise spectra for modes $a_T$ (green), $a_{N-1,N}$ (red, $\blacktriangle$ for $a_N$ and $\blacksquare$ for $a_{N-1}$), $a_0$ (black), and $a_{-1}$ (blue). The blueshifted mode is a coherent state that is approximately decoupled from the nonlinear interactions due to its low $Q$ factor. The pump and IR cascading orders have low frequency noise that lies far below the reference coherent state defined by a state with identical decay channel but no nonlinear coupling. However, these modes feature strong GHz relaxation oscillations (ROs). Multiple relaxation oscillation peaks (around the nonlinear rate $|\kappa a_T|$) are present due to the TWM processes occurring in the multi-resonant cavity. 
    In these simulations, the pump and seed wavelengths are $\lambda_{0,1}=1064,1068$ nm (so that $\omega_T=2\pi\cdot 1.06$ THz). $N=9$ cascading orders are simulated, along with two low $Q$ ``padding modes'' on either side of the frequency space cavity. $Q$ factors used are: $Q_r=10^7$ (redshifted modes in frequency space cavity), $Q_0=Q_N=10^5$ (frequency mirrors), $Q_b=10^2$ (blueshifted modes), and $Q_T=10^4$ (THz idler mode). The nonlinear strength is $\kappa=4.70$ s$^{-1}$ and the input pump and seed powers are $|s_{0,1}|^2=1$ MW.}
    \label{fig:fig2}
\end{figure}


Using the formalism described in Sec. \ref{sec:theory}, we can compute the Fano factor noise spectrum for an arbitrary mode $a_n$ as $\langle \delta n_n^\dagger (\omega)\delta n_n(\omega)\rangle/n_n=\langle \delta p_n^\dagger(\omega) \delta p_n(\omega)\rangle$ (where $n_n=a_n^\dagger a_n,\delta n_n=a_n\delta a_n^\dagger + a_n^*\delta a_n$ denote the intracavity photon number and its fluctuations for mode $a_n$). In Fig. \ref{fig:fig2}c, this noise spectrum is plotted for several frequency modes. The low $Q$ blueshifted modes do not have strong nonlinear coupling with other modes in the system and are governed by driven-dissipative dynamics, generating an intracavity coherent state (blue curve). By contrast, the idler mode (green), cascading infrared orders (red), and pump mode (black) undergo strong nonlinear interactions that dominate their dynamics. This results in strong gigahertz relaxation oscillations (on the order of the characteristic nonlinear rate $|\kappa a_T|$). Several relaxation oscillation peaks are present due to the strong nonlinear coupling between multiple modes within the cavity.  

In Eq. \ref{eq:inout}, at frequencies much smaller than the cavity bandwidth, the inverse fluctuation matrix has entries (for the cascading IR orders) governed by the smallest timescale in the system, in our case $M^{-1}_{n,k} \sim 1/\mathrm{max}(\kappa a_T, \gamma_0)$, and the noise for these modes is dominated by the leakiest (lowest $Q$) mode, which is generally the pump mode $a_0$. Thus, the low frequency noise for the cascading orders scales as $\langle \delta p^\dagger_n(0) \delta p_n(0) \rangle \sim \mathcal{O}(\gamma_0/\mathrm{max}(\kappa a_T, \gamma_0)^2)$. In Fig. \ref{fig:fig2}d, we show how, when compared to the noise of a ``reference coherent state'' with equivalent loss but no nonlinear coupling, the low frequency noise is of order $\frac{\gamma_0\gamma_n}{\mathrm{max}(\kappa a_T, \gamma_0)^2}\ll 1$ times that of the coherent state. Thus, the intracavity low-frequency noise reduction relative to the aforementioned coherent state is enhanced by maximizing the nonlinear rate $\kappa a_T$ while making the $Q$ factor for all cascading modes large, so that $\gamma_n^2|M^{-1}_{n,k}|^2\ll 1.$
This low-frequency noise reduction can be interpreted as the enhancement of noiseless nonlinear processes relative to dissipative outcoupling which, as we will see, permits output amplitude noise squeezing.

Finally, we note that in the systems we have examined, integrated intracavity noise appears to remain at the shot noise limit due to high frequency relaxation oscillations (ROs), resulting in multimode intracavity coherent states. System configurations that damp ROs for some (or all) modes and thus permit intracavity squeezing may exist, such as systems with saturable absorbers or other nonlinear losses \cite{pontula2022strong, rivera2023creating}. Nonlinear dissipation has been proposed as a method to generate strong single mode intracavity squeezing, and its multimode extension should be investigated. 


\section{Output noise squeezing}
\label{sec:output}

\begin{figure*}[t]
    \includegraphics[scale=0.8]{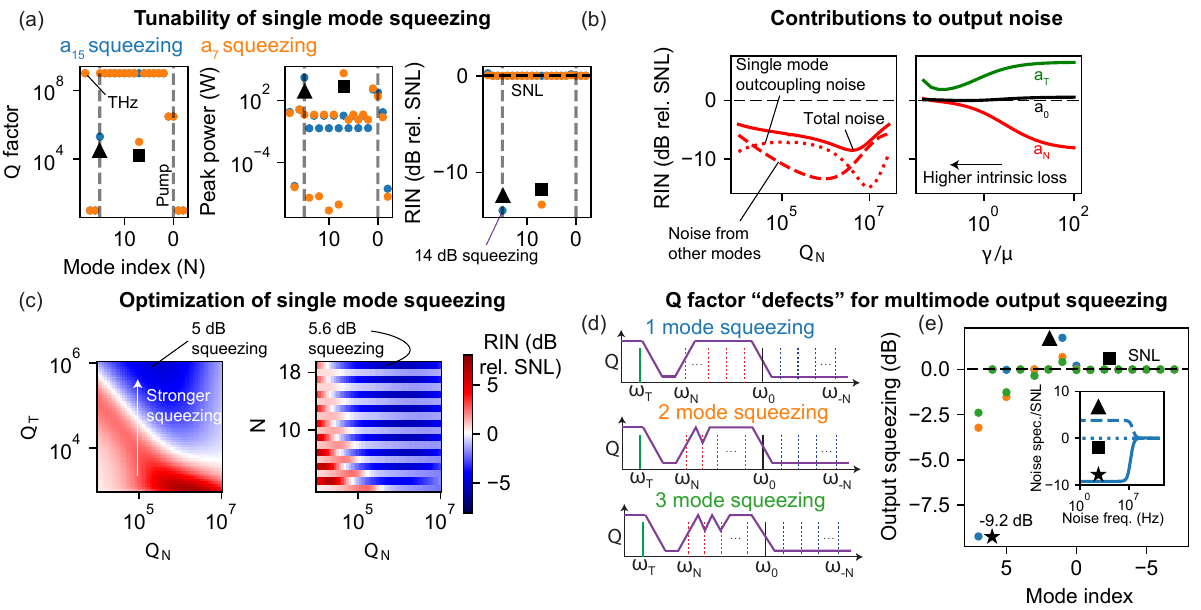}
        \caption{\textbf{Single and multimode output noise squeezing.} (a) Using a similar cavity design to that presented in Fig. \ref{fig:fig2}, single frequency modes in the synthetic frequency dimension can be squeezed in output noise. Here, we simulate $N=15$ cascading orders, and the mode with low $Q$ is squeezed. Dashed lines demarcate the boundaries of the frequency space cavity. We plot the steady state output power in the modes; due to their low $Q$, the squeezed modes have the highest power outside the cavity. The remaining IR modes show an interference pattern in steady state power characteristic of the Bloch interference phenomenon in Fig. \ref{fig:fig2}. The terahertz mode $a_T$ also has high power, while modes outside the frequency space cavity are very leaky and negligibly occupied in the steady state. The modes that are designed to have low $Q$ are the only ones to show significant departure from the shot noise limit (SNL), demonstrating intensity noise squeezing exceeding 10 dB. In these simulations, $Q_0=3\times 10^6,Q_N=2\times 10^5,Q_r=Q_T=10^9$ and $\kappa=14.1$ s$^{-1}$. (b), (c) The single mode output squeezing (here for mode $a_N$) can be maximized by optimizing multiple parameters simultaneously. Here, we show that higher $Q_T$ and lower intrinsic loss generates stronger squeezing due to stronger nonlinear energy flow. A larger number of modes ($N$) can also help increase squeezing, though too many modes can make the noise contribution from modes $a_{k\ne N}$ significant (red dashed curve in (c)). Lastly, an optimal $Q_N$ exists (with all parameters held equal) to maximize squeezing. Roughly, this $Q_N$ maximizes destructive interference with vacuum shot noise as per condition (1) (red dotted curve in (c)). In (b), (c), $Q_r=10^8$ and $\mu/\gamma$ denotes the ratio of intrinsic loss to the outcoupling rate. (d), (e) By shaping the $Q$ factor profile of the multimode cavity, specifically by introducing multiple $Q$ factor ``defects,'' output squeezing can be obtained for multiple frequency modes. Here, $N=7$ modes are simulated. The bandwidth for squeezing in the inset is around 100 MHz, but can be optimized to $>1$ GHz by enabling stronger nonlinear rates.}
        \label{fig:fig3}
\end{figure*}

\subsection{Single mode squeezing}

We now describe how strong output amplitude noise squeezing can be generated using cascaded nonlinear interactions in a multimode cavity. Notice from Eq. \ref{eq:inout} that there should be strong destructive interference between the intracavity fluctuations of mode $a_n$ and far-field vacuum fluctuations $s_n$ for the same mode (Fig. \ref{fig:fig1}b). At the same time, the noise contributions from all other modes should be minimized. Specifically, the conditions to maximize output squeezing read
\begin{align*}
    &(1) \quad [1-2\gamma_n \mathrm{Re}(M_{n,n}^{-1}(\omega))]^2\ll 1 \\
    &(2) \quad 4\gamma_n\gamma_k |(M_{n,k}^{-1}(\omega))|^2 \ll 1, k\ne n,
\end{align*}
where $M^{-1}_{n,k}$ is an element of the inverse fluctuation matrix that denotes the contribution of fluctuations in mode $a_k$ to the (intracavity) amplitude noise of mode $a_n$ and $\gamma_n$ denotes the outcoupling rate for mode $a_n$. These conditions are satisfied when the mode to be squeezed has a decay rate $\gamma_n$ on the order of the nonlinear rate $\kappa a_T$, while the other modes with non-negligible steady-state amplitude are of higher $Q$ factor (and other modes with low $Q$ factor are negligibly occupied). To see this, notice that for an ideal driven-dissipative state at zero noise frequency, $\mathrm{Re}(M_{n,n}^{-1}(0))=1/\gamma_n$ so the self-induced noise from condition (1) is at the shot noise limit, while for $\mathrm{Re}(M_{n,n}^{-1}(0))=1/(2\gamma_n)$ perfect destructive interference in condition (1) is achieved (zero self-induced noise in the output). In our system, this can be tuned by the ratio $\gamma_n/\kappa a_T$. When all other modes in the frequency comb have high $Q$ factor, the additive effect on noise in the outcoupled field due to other modes is minimal (condition 2), and squeezing can be observed.

Conditions 1 and 2 determine which discrete frequency modes can be squeezed. It is not possible for the high $Q$ intermediate cascading orders to be squeezed since condition 1 is violated. The external cavity noise for these modes is dominated by external vacuum shot noise. However, mode $a_N$, for example, terminates the frequency comb and thus has a larger outcoupling (and lower loaded $Q$ factor). When $\gamma_N=\mathcal{O}(\kappa a_T)$, destructive interference of the intracavity amplitude fluctuations with the external vacuum shot noise can occur. Condition 1 can be satisfied through optimizing $M_{N,N}^{-1}$, which in turn can be controlled by the $Q$ factor profile of the cavity. When all other cascading orders (including the pump and seed) are of higher $Q$ factor, both conditions 1 and 2 can be satisfied, yielding strong output amplitude noise squeezing for $a_N$ that exceeds 10 dB over nearly gigahertz bandwidths, as described below. 

In Fig. \ref{fig:fig3}a, we consider a one-sided comb with $N=15$ cascading orders where we selectively squeeze a terminal mode ($\lambda_{15}=1127$ nm) or a mode lying inside the frequency cavity ($\lambda_{9}=1101$ nm) by creating a low $Q$ factor defect for the squeezed mode in the otherwise high $Q$ factor frequency cavity (left panel of Fig. \ref{fig:fig3}a). In the center panel, notice that (1) the nonzero reflectivity of frequency mirror $a_N$ generates Bloch interference in the mean field and (2) the low $Q$ factor for the squeezed mode guarantees its large outcoupled power. Satisfying condition 1 (due to the low $Q$ factor of the squeezed mode) and condition 2 (due to the high $Q$ factor of all other coupled modes) generates strong single mode squeezing in low frequency output amplitude noise over 10 dB below the shot noise limit, as seen in the right panel.

In Fig. \ref{fig:fig3}b (first panel), we show the contribution of conditions (1) and (2) to the output noise in the terminal mode $a_N$. The total output noise is minimized when the sum of the two contributions is minimized. We also show the contribution of intrinsic loss to output squeezing (second panel), which reveals that strong squeezing in $a_N$ persists even when intrinsic loss is around $10\%$ of the outcoupling rate, i.e. $\gamma_n/\mu_n\approx 10$ (the same ratio of intrinsic loss to outcoupling is used for all modes). 

In Fig. \ref{fig:fig3}c, we examine how various setup parameters shape single mode squeezing in the terminal mode $a_N$. In the first panel, we see that stronger squeezing is achieved for higher $Q_T$. In addition to enhancing cycling of the terahertz idler photon within the cavity (which strengthens nonlinear energy flow), a higher $Q_T$ reduces the effect of coupling of fluctuations in $a_T$ to output noise in $a_N$. As expected from condition (1), with fixed $Q_T$, we see there is an optimal $Q_N$ to generate strongest squeezing. When $Q_N$ is too low or too high, destructive interference with the external vacuum field is ineffective. In the second panel, we sweep over $Q_N$ and the number of cascading orders $N$. The most distinctive feature is the weakened squeezing for even $N$. This occurs due to the effect of Bloch interference, specifically the pump mode. When $N$ is odd, the low $Q$ factor end of the frequency space chain $a_0,a_{-1}$ are negligibly occupied, so the squeezing for $a_N$ is strong. When $N$ is even, the noise contribution from $a_{0},a_{-1}$ is significant since they are non-negligibly occupied, so the squeezing in $a_N$ is less due to noise coupling to $a_{0},a_{-1}$. We also notice a tendency towards stronger squeezing for longer combs (larger $N$). This appears to be because of an inverse scaling with $N$ of the coupling of the idler mode fluctuations to the output noise in $a_N$, due to an enhancement in the effective nonlinear rate relative to dissipation rates. We have found that when condition (1) is fully satisfied, the output noise (relative to the SNL) in $a_N$ goes as $\frac{\gamma_T|a_T|^2}{N^2\gamma_N|a_N|^2}$. This holds as long as the idler mode is the dominant source of (coupled) noise and may break down for very large $N$ when the additive contribution of the noise coupling from the high $Q$ infrared cascading orders becomes significant.     


\subsection{Multimode squeezing}

By introducing multiple low $Q$ factor ``defects'' into the chain of cascading orders, multiple frequency modes can be squeezed in output amplitude noise. When Bloch interference is present, modes with opposite parity to mode $a_N$ have significantly damped low-frequency intracavity noise, so the destructive interference with external vacuum shot noise is ineffective. Therefore, modes $a_{N}, a_{N-2},...$ with the same parity as $a_N$ (high steady-state amplitude, high intracavity noise branch) are more strongly squeezed. As the number of modes we would like to squeeze increases, the degree of squeezing decreases. Suppose we aim to squeeze $m$ modes by introducing $m$ identical ``defects'' ($Q$ factor dips). In the ideal case, the noise is dominated by these low $Q$ modes. Let $x$ denote the quantity $2\gamma_n M_{n,n}^{-1}$ for one such squeezed mode. Then, for equal squeezing in all modes, we compute
\begin{align}
 \min_{x}\left[(1-x)^2+(m-1)x^2\right] = 1-\frac{1}{m}.   
\end{align}
Notice that this gives the 3 dB single beam output squeezing limit of the signal and idler beams when $m=2$ \cite{fabre1989noise}. As $m$ grows large, all modes approach the shot noise limit.

In Fig. \ref{fig:fig3}d, e, we show how introducing 1 ($a_N$ squeezed), 2 ($a_N,a_{N-2}$ squeezed), and 3 ($a_N,a_{N-2}, a_{N-4}$ squeezed) defects generates single and multimode output amplitude noise squeezing. By further tuning the $Q$ factors of the squeezed modes, it may be possible to control the ``distribution of squeezing'' over the squeezed modes (e.g., in the trivial case, only one low $Q$ factor defect corresponds to single mode squeezing). As expected from the previous discussion, the squeezing for multiple modes weakens. The inset shows that the bandwidth over which squeezing occurs is similar to that for intracavity squeezing and limited by the onset of relaxation oscillations. The amplitude noise returns to shot noise level around 100 MHz-1 GHz. This bandwidth is limited by the onset of intracavity relaxation oscillations (i.e. the nonlinear rate in our system). Thus, strong nonlinear interactions can in principle reach GHz-surpassing bandwidths.

\section{Multimode twin beam quantum correlations}
\label{sec:correl}

\begin{figure}
    \centering
    \includegraphics[scale=0.8]{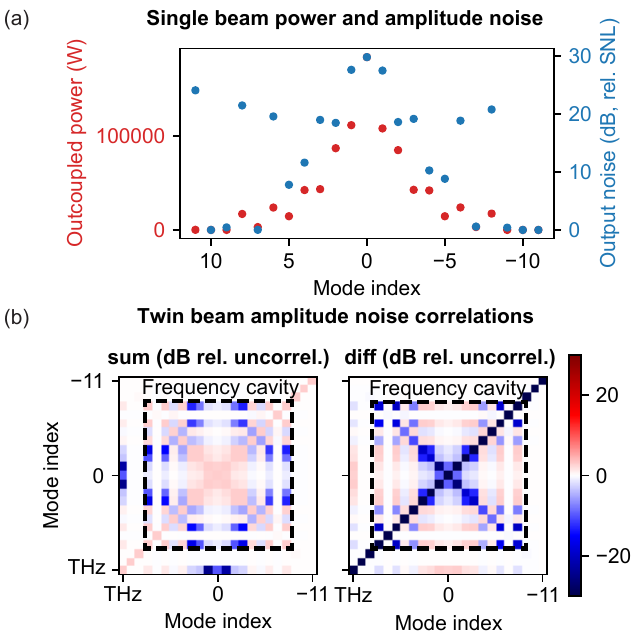}
    \caption{\textbf{Twin beam correlations.} (a) Single beam outcoupled power and (DC) output noise. (b) Twin beam intensity sum and difference fluctuations $\langle \delta n_i \pm \delta n_j\rangle$ normalized to the uncorrelated twin beam noise. Despite certain modes being strongly antisqueezed in (individual) output amplitude noise, strong correlations between multiple pairs of modes significantly reduce the twin beam noise. These correlations span the dimension of the frequency cavity and may point towards the possibility of long-range entanglement in a synthetic frequency dimension. Simulation parameters are $Q_0=Q_N=Q_r=Q_b=3\times 10^6,Q_T=10^5,\kappa = 3\times 10^{-4}$ J$^{-1/2}$, and $|s_0|^2=1$ MW, $|s_1|^2=100$ W. All noises are computed at noise frequencies much lower than the cavity bandwidth.}
    \label{fig:fig4}
\end{figure}

In this section, we examine quantum noise correlations between discrete frequency modes in our system. In the case of a single ideal TWM process, it is well known that the amplitude sum of the signal and idler is noiseless and signal and idler photons are entangled \cite{boyd2008nonlinear}. When cascaded nonlinear processes are present, this entanglement is now distributed over many modes. In Fig. \ref{fig:fig4}, we consider a two-sided comb (in contrast to the one-sided comb design for squeezing above; see Appendix for $Q$ factor profile) and plot the low frequency noise in the output intensity sum $S_{ij}=\sum_{k,k' \in \{i,j\}} \langle \delta n_k^\dagger \delta n_{k'}\rangle$ and difference  $D_{ij}=\sum_{k,k' \in \{i,j\}} (-1)^{1+\delta_{k,k'}}\langle \delta n_k^\dagger \delta n_{k'}\rangle$, where $n_k=a^\dagger_ka_k$ is the number operator for mode $k$ (further details on the calculation are provided in the Appendix). These twin beam noises are normalized to the uncorrelated twin beam noise $U_{ij}=\sum_{k \in \{i,j\}} \langle \delta n_k^\dagger \delta n_k\rangle$. Normalization by $U_{ij}$ shows that twin beam squeezing emerges due to \textit{correlations} between the two modes rather than single beam squeezing.

We can compare the twin beam noise to the noise in a single mode. In Fig. \ref{fig:fig4}a, we plot the outcoupled power and low frequency output amplitude noise for individual modes. Certain modes are near the SNL, while others are strongly anti-squeezed. Fig. \ref{fig:fig4}b demonstrates that twin beam noise can be reduced by orders of magnitude relative to single beam noise. This strong squeezing relative to single beam noise generally occurs when the two modes have comparable individual noise, as this permits stronger destructive interference in the amplitude fluctuations. We see that strong correlations can occur within the frequency cavity and with the common terahertz idler mode, resulting in squeezing over 20 dB relative to the uncorrelated twin beam noise. In contrast to the correlations for a single TWM process, the correlations in our system can be much longer range, spanning the dimension of the frequency cavity. We also point out the twin beam squeezing in the noise of the amplitude difference $p_n-p_{-n}$. This is reminiscent of the strong squeezing reported in the supermodes of soliton microcombs, $p_n\pm p_{-n}$ \cite{guidry2023multimode}. We conclude by noting that recent experiments have reported strong noise correlations for multiple wavelength pairs in the continuous spectrum generated by a nonlinear fiber, providing impetus for realizing multimode quantum states over discrete frequency modes \cite{uddin2023ab}.     

%
%

\section{Discussion}

In this paper, we demonstrated bright single- and multimode squeezing using cascaded three wave mixing processes in a nonlinear cavity. Our work constitutes a distinct paradigm shift relative to most previous works that have focused on below-threshold parametric squeezing in the single- and multimode regimes. Furthermore, we have shown the existence of quantum correlations between multiple pairs of frequency modes, extending the concept of twin beam squeezing that is well-known for single parametric downconversion processes. In this section, we provide an outlook on this work from both a theoretical and experimental perspective.

We have noted previously that intracavity squeezed state generation in the proposed system is difficult given the existence of high frequency relaxation oscillations. However, the generation of multimode intracavity bright squeezed states could enable a new regime of cavity QED experiments \cite{qin2018exponentially, zeytinouglu2017engineering, burd2021quantum, leroux2018enhancing}. For example, single mode cavity QED is generally limited to global interactions well-described by mean field theory, whereas multimode cavity QED may permit tunable local couplings that can elucidate beyond-mean-field physics \cite{vaidya2018tunable}. Thus, mechanisms for the suppression of relaxation oscillations should be investigated, such as recent theoretical work on the application of nonlinear dissipation to single mode intracavity squeezing in lasers \cite{pontula2022strong, rivera2023creating}. 

Exciting topological phenomena have been studied in synthetic dimensions in photonics, and our work suggests a platform for studying the intersection of topology and nonlinear quantum optics in a synthetic frequency dimension \cite{yuan2018synthetic, lustig2021topological}. For example, recent work has explored the use of external amplitude and phase modulation in ring resonators and coupled OPOs to generate non-Hermitian tight-binding coupling between resonant frequency modes \cite{roy2021nondissipative, wang2021generating, yuan2021synthetic}. Our system offers the opportunity to tune both nonlinear coupling (as we have done here) and non-Hermitian modulation, shaping energy propagation between discrete frequency modes. This could unlock novel topological phenomena such as skin effects in a synthetic frequency dimension and topologically-protected quantum optical states.

Recent work with electro-optic modulated thin-film lithium niobate microresonators has revealed the potential of using interference between Bloch modes to tailor the flow of light in the synthetic frequency dimension, creating, for example, trapped states \cite{hu2022mirror}. Applying similar techniques to our system could allow creation of squeezed frequency-space solitons and other more exotic classical and quantum states of light. Additionally, we anticipate that squeezing in these Bloch modes or quasi-Bloch modes that diagonalize the nonlinear Hamiltonian could be even larger than the squeezing for individual frequency modes described here, inspired by recent proposals to generate output squeezing exceeding 15 dB in supermodes of a soliton microcomb system \cite{guidry2023multimode}. 

We now comment on experimental platforms that may realize the effects described in this paper. The important criterion to generate squeezing and strong long-range correlations is a strong enhancement of the nonlinear coupling relative to dissipation in the system, which requires (1) high pump and seed power, (2) a strongly resonant nonlinear multimode cavity, and (3) a method to tune the dissipation ($Q$ factor) for different frequency modes. In addition to free space optical parametric oscillators (OPOs), on chip OPOs may offer a platform to realize the effects described here with compact form factor \cite{ledezma2023octave}. Recent advances in the integration of lithium niobate photonics with ultra high $Q$ whispering gallery mode resonators \cite{lin2015fabrication, hao2018periodically}, for instance, may provide the necessary elements to generate cascading nonlinear processes, though intrinsic losses (particularly at the idler frequency) will need to be minimized. 

The effects we have described here do not depend on specific spectral ranges for the pump, signal, and idler modes. Depending on the platform and material, a higher frequency mid-IR idler mode could be used instead, potentially with lower losses at the expense of a shorter comb. Furthermore, an essential part of our approach is $Q$ factor engineering, which allows one to tune the length of the frequency comb, relative amplitudes of the different modes, and which modes are squeezed/correlated. Experimentally, this $Q$ factor engineering can be achieved by using photonic crystals that provide frequency-tunable filters for coupling into and out of the cavity.

Our work establishes the mechanism of cascaded nonlinear optical processes as a method to generate frequency combs that exhibit bright squeezing and quantum correlations over a broad (and tunable) spectral range. We envision future application of the concepts described here to tunable squeezed light sources, multimode entanglement for sensing and quantum computing protocols, and much more.   

\section{Acknowledgements} We acknowledge useful discussions with Nicholas Rivera, Jamison Sloan, and Shiekh Zia Uddin. S.P. acknowledges the financial support of the Hertz Fellowship Program and NSF Graduate Research Fellowship Program. Y. S. acknowledges support from the Swiss National Science Foundation (SNSF) through the Early Postdoc Mobility Fellowship No. P2EZP2188091. C. R.-C. is supported by a Stanford Science Fellowship. This material is based upon work supported in part by the Air Force Office of Scientific Research under the award number FA9550-20-1-0115; the work is also supported in part by the U. S. Army Research Office through the Institute for Soldier Nanotechnologies at MIT, under Collaborative Agreement Number W911NF-23-2-0121. We also acknowledge support of Parviz Tayebati.

\bibliography{fock}

\end{document}